\documentclass[preprint]{aastex}

\usepackage{epsf}

\begin{document}

\title{Three-dimensional Simulations of the Parker Instability in
a Uniformly-rotating Disk}
\author{Jongsoo Kim\altaffilmark{1}, 
        Dongsu Ryu\altaffilmark{2}
        and T. W. Jones\altaffilmark{3}}

\altaffiltext{1}
{Korea Astronomy Observatory, 61-1, Hwaam-Dong, Yusong-Ku, Taejon 305-348, 
Korea and National Center for Supercomputing Applications, University of Illinois at Urbana-Champaign, 405 North Mathews Avenue, Urbana, IL 61801: jskim@kao.re.kr, jskim@ncsa.uiuc.edu}
\altaffiltext{2}
{Department of Astronomy \& Space Science, Chungnam National University,
Daejeon 305-764, Korea:\\ryu@canopus.chungnam.ac.kr}
\altaffiltext{3}
{Department of Astronomy, University of Minnesota, Minneapolis, MN 55455:\\
twj@mail.msi.umn.edu}

\begin{abstract}

We investigate the effects of rotation on the evolution of the Parker
instability by carrying out three-dimensional numerical 
simulations with an isothermal magnetohydrodynamic code.
These simulations extend our previous work on the nonlinear
evolution of the Parker instability by Kim et al. (1998).
The initial equilibrium system is composed of exponentially 
stratified gas and field (along the azimuthal 
direction) in a uniform gravity (along the downward vertical direction).  
The computational box, placed at the solar neighborhood, is set to rotate 
uniformly around the galactic center with a constant angular speed.
The instability has been initialized by random velocity perturbations.
In the linear stage, the evolution is not much different from
that without rotation and the mixed (undular + interchange) mode
regulates the system. The interchange
mode induces alternating dense and rarefied regions with small radial
wavelengths, while the undular mode bends the magnetic field lines in
the plane of azimuthal and vertical directions.
In the nonlinear stage, flow motion overall becomes chaotic as
in the case without rotation.  However, as the gas in higher
positions slides down along field lines forming supersonic flows,
the Coriolis force becomes important.
As oppositely directed flows fall into valleys along both sides
of magnetic field lines, they experience the Coriolis force toward
opposite directions, which twists magnetic field lines
there.  Hence, we suggest that the Coriolis force plays a role
in randomizing magnetic field.
The three-dimensional density structure formed by 
the instability is still sheet-like with the short dimension along the
radial direction, as in the case without rotation.  
However, the long dimension is now slightly tilted with respect to the
mean field direction. The shape of high density regions is a bit rounder.
The maximum enhancement factor of the vertical column density relative to 
its initial value is about 1.5, which is smaller than that in the case 
without rotation.  We conclude that uniform rotation doesn't
change our point of view that the Parker instability {\it alone} is not
a viable mechanism for the formation of giant molecular clouds. 

\end{abstract}


\section{INTRODUCTION}

The nature of the galactic magnetic field has been revealed from 
observations of pulsar dispersion measures, starlight polarization 
and synchrotron emission (Heiles 1996 and references therein).  It is 
composed of uniform and random components. Their estimated strengths 
at the solar neighborhood are of order a few $\mu$G.  The scale of
irregularities of the random component measured by Rand \& Kulkarni
(1989) is about $\sim 50$ pc.
The magnetic energy density is comparable to those
of other energy contents such as cosmic-rays, turbulence and gravity.
Therefore, the magnetic field forms one of the important constituents of
the interstellar medium (ISM) and plays a key role in its evolution.
Parker (1966) demonstrated that the ISM is unstable due to magnetic
buoyancy.  The instability, referred to as the {\it Parker instability},
is an example demonstrating that the magnetic field is important in the
dynamics of the ISM.

Two different approaches are employed to investigate the nature of
instabilities. One is the linear stability analysis, which enables us to
find unstable modes and estimate their growth rate.   
Parker (1966, 1967) himself used this approach to show the existence of
the  instability.  Later, other authors elaborated on the instability 
with the same approach. A partial list includes the studies on the effects
of galactic rotation (Shu 1974; Zweibel \& Kulsrud 1975; Foglizzo \&
Tagger 1994, 1995), magnetic micro-turbulence (Zweibel \& Kulsrud 1975),
a skewed configuration of magnetic field (Hanawa, Matsumoto, \& Shibata  1992), 
a non-uniform
nature of externally given gravity (Horiuchi et al. 1988; Giz \& Shu 1993;
Kim, Hong, \& Ryu 1997;  Kim \& Hong 1998), the galactic corona
(Kamaya et al. 1997), and multiple gas components (Kim et al. 2000).

According to the linear analysis, the instability is broken down into
two distinct modes: {\it undular} and {\it interchange}
(Hughes \& Cattaneo 1987; Kim et al. 1998).
The former, whose perturbations are defined in
the plane of gravitational field and unperturbed magnetic field, undulates
magnetic field lines and induces gas to slide down along the field lines
into magnetic valleys.  The latter interchange mode, whose perturbations
are defined in the plane perpendicular to the direction of unperturbed 
magnetic field, generates alternating dense and rarefied regions
along the radial direction perpendicular to both the gravity and
field directions and has a maximum growth rate at infinite radial wavenumber,
indicating it is a Rayleigh-Taylor type.  When perturbations are allowed
in all three spatial dimensions, the {\it mixed mode} of the undular and
interchange modes appears (Matsumoto et al. 1993), which has
a maximum growth rate at infinite radial wavenumber too.

The other approach is through numerical simulations, where the nonlinear
evolution of instabilities is studied.  Since the first
one-dimensional study by Baierlein (1983), many studies on the Parker
instability under different conditions have been done:
Matsumoto et al. (1990) and Matsumoto \& Shibata
(1992) explored the 
nature of the undular and mixed modes in accretion disk environments. 
Shibata et al. (1989a, 1989b), Kaisig et al. (1990) and Nozawa et al. 
(1992) considered the Parker instability as the 
driving mechanism for emergent flux tubes in the Sun, and tried to explain 
several features of the solar activity.  Basu, Mouschovias, \& Paleologou (1997) and
Kim et al. (1998) performed multi-dimensional simulations
of the instability in a thin gaseous disk under the influence of a
simplified uniform gravity and addressed the issue of the origin of the giant
molecular clouds (GMCs).
Santill\'an et al. (2000) simulated the evolution of the Parker instability
in a more realistic galactic disk with multiple gas components.

Especially, Kim et al. (1998) used high-resolution three-dimensional
simulations to follow the nonlinear evolution of the mixed mode.
They found that after linear growth perturbations become saturated in
the nonlinear stage while forming chaotic structures.  On top of that,
alternating regions of magnetic arches and valleys are developed.
The chaotic motions induce reconnection,
especially in the valley regions, which allows gas to cross field lines.
As a result, gas and magnetic field are redistributed into
a stable system in the relaxed stage.

In this paper we begin to investigate the effects of rotation on the nonlinear
Parker instability.
For simplicity, we consider {\it uniform rotation}.  Eventually, it is
necessary to address the effects of {\it differential rotation}.
But we will leave it for a future work.

It is known through linear analyses (Shu 1974; Zweibel \& Kulsrud 1975)
that uniform rotation reduces the growth rate of the undular mode. 
It is the Coriolis force that prevents the lateral motion and reduces
the growth rate.  However, as shown in \S 2, the Coriolis force affects
less the linear growth of the three-dimensional mixed mode. The reason is the
following:  The normal modes have small wavelengths
along the radial direction, and the most unstable mode has a vanishing
wavelength. Hence, the perturbed velocity in the linear stage is predominantly
along the vertical direction, the direction of the rotation vector, so it
does not interfere with the rotation.
But yet, from this it is not clear how the Coriolis force affects the nonlinear
evolution.  Related to that, Chou et al. (1997)
studied rotating flux tubes numerically, modeling the evolution
of the emergent magnetic flux sheet.  They demonstrated that the emergent flux
sheet becomes twisted due to the Coriolis force, and explained
the tilted active regions observed in the Sun in terms of twisted flux sheets.
They, however, chose perturbations with a 
pair of finite horizontal wavelengths that do not correspond to 
the mode of the maximum growth rate, and performed numerical experiments
with low resolution.  
Miller \& Stone (2000) studied in the environment of accretion disks
the effects of differential rotation.  They focused on the role of
the magneto-rotational instability, but also commented
on the role of the Parker instability too.
Here, we study the nonlinear evolution of the Parker instability
in the environment of the galactic disk, which is initiated by 
{\it random perturbations}, through {\it high-resolution},
{three-dimensional} numerical simulations of the Parker instability.
High-resolution and three-dimensions are
required to resolve the small-scale structures of the mixed mode.
Random perturbations are necessary to find the mode preferred
by the system.

The Parker instability has been thought to be a mechanism in forming
the GMCs in the Galaxy (Appenzeller 1974;
Mouschovias, Shu, \& Woodward 1974; Blitz \& Shu 1980; Shibata \& Matsumoto 1991;
Handa et al. 1992; Gomez de Castro \& Pudritz 1992).  The conjecture is
based on the results from two-dimensional linear analyses of the
undular mode that the growth time 
and wavelength of the most unstable mode are comparable to the life 
time and the separation of the GMCs.  However, it has been known from
earlier work by Parker (1967) that the 
instability initiated by three-dimensional perturbations develops
the mixed mode at its maximum growth rate with vanishing wavelength
along the radial direction.  In addition, Kim et al. (1998) showed 
in three-dimensional simulations that sheet-like structures with
smallest scale along the radial direction, whose shape is different from that
of observed GMCs, form at the developing stage of the instability and
persist during the nonlinear stage.
Furthermore, they showed that the enhancement factor of column density 
relative to its initial value is only $\sim2$, which is too small for
the GMC formation.  We will revisit the issue of the
GMC formation by the Parker instability with the results on the
effects of rotation on the instability.

The plan of the paper is as follows.  In \S 2, we
summarize the problem including numerical set-up and linear analysis.
In \S 3, results of numerical simulations are presented.
Summary and discussion follow in \S 4.

\section{PROBLEM}

\subsection{Numerical Setup}

In order to describe the effects of rotation
on the Parker instability at the Solar 
neighborhood, we introduce a local Cartesian coordinate system, which is 
rotating with the galactic angular velocity $\Omega \hat{z}$ at the Sun's
position, $R_0$.  In this local frame, the coordinates are defined as 
$x = (R-R_0)$, $y=R(\phi - \Omega t)$, and $z$, and their directions
are parallel to radial, azimuthal, and vertical directions of the
Galaxy, respectively.  Assuming a constant angular speed, the isothermal 
magnetohydrodynamic (MHD) equations with a downward uniform gravity 
($-g \hat{z}$) are:  
\begin{equation}
\frac{\partial \rho}{\partial t} + \nabla \cdot (\rho {\bf v}) = 0,
\end{equation}
\begin{equation}\label{momentum}
\frac{\partial}{\partial t}(\rho {\bf v}) 
+ \nabla \cdot \left( \rho {\bf v}{\bf v} + \rho a^2 {\bf I}
+ \frac{B^2}{8\pi}{\bf I} - \frac{{\bf B}{\bf B}}{4\pi} \right) =
-2 \rho \Omega \hat{z} \times {\bf v} - \rho g \hat{z},
\end{equation}
\begin{equation}
\frac{\partial {\bf B}}{\partial t} + 
\nabla \cdot ({\bf v}{\bf B} - {\bf B}{\bf v})=0,
\end{equation}
where $a$ is an isothermal sound speed and other notations have
their usual meanings. Note that we don't take into account the centrifugal
force in Eq. (\ref{momentum}), since the force is canceled out by 
the radial component of the gravitational force.
The above equations are the same as those solved in Kim et al. (1998)
except the Coriolis force term in the momentum equation.

As in Kim et al. (1998), the initial distributions of gas and 
magnetic field have been set up in magnetohydrostatic equilibrium.  
Dropping the terms containing
time derivative and velocity in Eq. (\ref{momentum}) and 
assuming i) uni-directional horizontal magnetic field
($B_0[z]\hat{y}$), and ii) a constant ratio of magnetic to gas 
pressures $\alpha = B_0^2[z]/(8\pi \rho_0[z]a^2$), then 
density and magnetic pressure
are described by an exponential function,
\begin{equation}\label{IC}
\frac{\rho_0(z)}{\rho_0(0)} = 
\frac{B_0^2(z)}{B_0^2(0)} =
\exp \left( - \frac{|z|}{H} \right),
\end{equation}
where the scale height of the gas is defined as $H \equiv (1+\alpha)a^2/g$.

A computational cube with $0 \le x,y,z \le 12H$ is placed at the solar 
neighborhood, where $12H$ is the azimuthal wavelength of the maximum linear 
growth for the undular mode (see the next subsection).    
To initiate the instability, random velocity perturbations are added.  
The standard deviation of each velocity component is set to be $10^{-4}a$.
Boundaries are periodic along the radial and azimuthal ($x$ and $y$)
directions and reflecting along the vertical ($z$) direction.
As the result, fluid quantities are conserved and expulsion of
magnetic flux through the upper boundary is prohibited.

The above isothermal MHD equations are solved by an
MHD code described in Kim et al. (1999), which is the isothermal version
of the MHD code based on the explicit, finite-difference Total Variation
Diminishing (TVD) scheme (Ryu \& Jones 1995; Ryu, Jones, \& Frank 1995).
Three-dimensional simulations have been performed for the cases with
and without rotation, using different resolutions.
Model parameters of the simulations are summarized in Table 1.
Model $1-3$ are the cases without rotation.
The highest resolution case, Model 3,
is the same one as that reported in Kim et al. (1998)
except that Model 3 extends to a longer end time, $t_{end}$.
We note that Model 3 does not simply extend that of Kim et al. (1998),
but has been re-calculated with a different set of initial random numbers.
The case with rotation, Model 4, has the angular velocity $\Omega=1/2 (a/H)$,
which is consistent with the galactic circular speed
(27 km s$^{-1}$ kpc$^{-1}$)
based on Hipparcos data (Feast \& Whitelook 1997).
All models are assumed to have $\alpha=1$, so that magnetic pressure
is initially the same as gas pressure.

Physical quantities of length, speed, density and magnetic field are given
in units of the density scale height, $H$, the isothermal sound speed, $a$,
the initial density at $z=0$, $\rho_0(0)$, and the initial field strength at
$z=0$, $B_0(0)$, respectively.  For the typical values of 
$a=6.4$ km s$^{-1}$ and $H=160$ pc (Falgarone \& Lequeux 1973), 
the resulting time unit, $H/a$, becomes $2.5\times10^7$ years.

It is well known that the real interstellar medium is composed of
multiple components with various scale heights.  Some of them have the 
scale heights different from our choice, 160 pc (see Eq. [5] in
Kim et al. 2000). So our single gas component model under uniform
gravity would be too simplistic.  However, it is the model which has
been most extensively used since the pioneering work done by
Parker (1966).  Especially, the effect of uniform rotation on the
instability was already investigated through linear analyses using 
the same model (see next subsection).  Choosing the well-studied
model would enable us to single out the consequences of rotation on
the Parker instability in the numerical simulations, which is the
primary goal of this paper. We note that two-dimensional numerical
simulations of the Parker instability under a more realistic
Galactic disk model were performed in Santill\'an et al. (2000) as
noted in the Introduction.

\subsection{Stability Analyses}

Criteria for the Parker instability are given in terms of an
effective adiabatic index.  If the adiabatic index is less than
a  critical value, the above system becomes unstable
(Parker 1966, 1967; Hughes \& Cattaneo 1987; Matsumoto et al. 1993).
The critical adiabatic index for the undular mode is
$\gamma_{\rm u}=(1+\alpha)^2/(1+\frac{3}{2}\alpha)$, while that
for the interchange mode is $\gamma_{\rm i}=1-\alpha$.  The
critical adiabatic index for the mixed mode is $\gamma_m=1+\alpha$.
Our consideration of isothermality corresponds to $\gamma=1$ and
the models have $\alpha=1$.  So the initial system is unstable against
the undular mode as well as the mixed mode, but stable to the
interchange mode.

Dispersion relations from linear stability analyses provide
the growth rate and wavelength of
unstable modes and help set the size of the computational cube.  
In Fig. 1, the dispersion relations for the
exponentially-stratified magnetized disk are reproduced
(see, Parker 1966, 1967; Shu 1974; Zweibel \& Kulsrud 1975).
The solid line represents the unstable undular mode,
which is initiated by two-dimensional ($y$ and $z$) perturbations, without 
rotation. Perturbations undulate azimuthal field lines and induces 
gas to slide down along the field lines into magnetic valleys.    
The undular mode has a preferred wavelength ($\sim 12H$) along the azimuthal 
direction, which has been set as the size of our computational cube.      
Inclusion of rigid-body rotation in the undular mode reduces the growth
rate as well as the wavelength of the most unstable mode
(see the dashed line), because lateral 
motion along the field lines is hampered by the Coriolis force.
When additional radial perturbations are applied, both
the undular and interchange mode work together,
and the resulting mixed (undular + interchange) mode
becomes more unstable than the undular mode alone.  
In the extreme case with infinite radial wavenumber, the growth rates
do not depend on angular speed (see the dotted line).  
This is because the perturbations with large radial wavenumbers
generate mostly vertical motions, which do not interfere with
the Coriolis force.  With infinite radial wavenumber,
the growth rate of the most unstable
mode is $\sim0.41$ and its wavelength is $\sim 11H$.
The dot-dashed line shows the dispersion relation with radial wavelength
equal to $\pi$, about a quarter of the box size.
The maximum growth rate $\sim0.39$,
in this case, is slightly smaller than that for vanishing radial wavelength,
but the wavelength of the most unstable mode is similar in the two
situations. Our simulations
correspond to conditions somewhat between the dotted and dot-dashed lines.

It is interesting to note that while the pure interchange mode is
stable,  we see the characteristics of the interchange mode  
in the dispersion relation of the mixed mode.  The reason is the
following.  When three-dimensional
perturbations are applied to the initial system, it quickly adjusts
itself by forming alternating regions of compression and rarefaction 
along the radial direction.  Both alternating regions are now subject 
to the undular instability.  For smaller wavenumbers, the undular
instability in each rarefied region develops faster with the help of 
buoyancy due to the neighboring dense regions.  That is why
the fastest growing mode has $k_x=\infty$.

\section{RESULTS}

\subsection{Effect of Resolution on Radial Wavenumber} 

The dispersion relations in Fig. 1 have shown that the most unstable 
mode has vanishing wavelengths along the radial direction.  This has
guided us to use the highest resolution (using up to $256^3$ grid zones)
achievable with our computing resources.
Even with the highest resolution, it is however not possible
to resolve all the small scale structures that are expected to
form in the simulations.  Bearing this fact in mind, we compare in Fig. 2
isodensity surfaces at the same epoch ($t=35$) but with different
resolutions using $64^3$, $128^3$, and $256^3$ zones
(Model 1, 2 and 3).  The computational boxes are   
oriented in such a way that the radial ($x$), the azimuthal ($y$),
and the vertical ($z$) directions are from left to right, from near to
far, and from bottom to top, respectively.  A gray surface with equal density
$\rho_0(z=4)$ is included in each box, which was initially a flat surface 
at $z=4$.  The first impression is that the density structure
in the low-resolution ($64^3$ zones) simulation looks very
different from that in the high-resolution ($256^3$ zones).  Similarly,
it looks like large-scale, ordered structures are formed in the low-resolution.
On the other hand, small-scale, chaotic structures are clear in
the high-resolution, which are expected from linear stability analyses.
This justifies the usage of resources to perform high-resolution
simulations.  In what follows, we discuss only the results
of high-resolution simulations, Model 3 and 4.

Since the instability has been initiated by random perturbations,
no preferred wavelength has been introduced.  But the upper and lower
limits of available wavelengths are set by the finite size of the
computational box, which is $12H$, and the numerical resolution
of the code, respectively.  In general, the number of grid zones required to
resolve one wavelength is different from problem to problem and from code
to code.  It is our experience that our code needs $\sim8$ grid zones
along one direction to cover one wavelength in the current simulations
without excessive numerical dissipation.
With 256 zones, the smallest wavelength that can be resolved would be
$\sim3/8H$, and $\sim32$ wavy structures are accommodated along one
direction.  

\subsection{Time Evolution of Root-Mean-Square Velocities}

Fig. 3 plots the time evolution of the rms velocity (volume averaged, not
mass averaged) in the simulations for the case without rotation ($\Omega=0$,
left panel) and with rotation ($\Omega=1/2$, right panel).
The $x$, $y$, and $z$ components are drawn with
dotted, short dash, and long dash lines, respectively.  The solid line
corresponds to the maximum linear growth rate, 0.41.
During the initial transient stage ($0 \leq t \lesssim 25$), random
perturbations adjust themselves by generating one-dimensional
magnetosonic waves, which move vertically, as pointed out in Kim et al.
(1998) and Santill\'an et al. (2000).
On top of the magnetosonic waves, there are slow MHD waves that
eventually develop into the unstable mode (Shu 1974; Kim et al. 1997).
As the slow MHD waves grow, the fluctuating and propagation features
of the magnetosonic waves gradually diminish.

As in Kim et al. (1998), the evolution can be divided into three
stages: {\it linear}, {\it nonlinear}, and {\it relaxed}.  The linear
stage lasts until $t \simeq 35$ for $\Omega=0$ and $t \simeq 32$
for $\Omega=1/2$.  In both cases, the slope of linear growth is close
to 0.41.  This demonstrates that numerical resolution is fine
enough to follow the modes that have small radial wavelengths.
At the end of the linear stage, the rms of $v_y$ is saturated
near $0.7$, which is comparable to the isothermal
sound speed.  This implies that the linear approximation
goes well beyond its conventional limit.

After saturation, the evolution enters the nonlinear stage.
As pointed out in Kim et al. (1998), due to the manifestation of the
interchange mode, the structures formed in the nonlinear stage are
characterized as chaotic (see \S 3.4).  On top
of the chaotic structures, the undular mode causes magnetic field lines
to drop at valley regions and rise at arch regions.
Along those undulated field lines, gas supersonically falls
toward valley bottoms and is stopped by shock waves.  The gas ends up
forming condensations at the valley bottoms.  In the case without
rotation, the condensations become 
over-compressed due to the continual falling of the gas.
This is apparent in the left panel of Fig. 3 for $\Omega=0$
around $t \simeq 55$, when the rms of $v_z$ shows its local minimum.
After that, the increased pressure pushes the valley matter back to
the upper regions.  This is recurrent, but with the rms of $v_z$
decreasing.  At the same time, reconnection, enhanced by chaotic
motions, allows re-distribution of mass and magnetic field and the system
evolves into a new equilibrium state described in Kim et al. (1998)
that is stable against the instability.
For this reason $t \gtrsim 55$ is labeled as the
relaxed stage.  However, in the case with rotation,
the in-falling gas speed is reduced by the Coriolis force, and so
the signature of a local minima is not quite so dramatic.  Hence,
distinguishing the relaxed stage from the nonlinear stage is less obvious.

A close examination of Fig. 3 reveals the following further
differences between the two cases.
i) The rms of $v_y$ reaches its peak value
0.78 at $t \simeq 40$ for $\Omega=0$, while 
the peak value reaches 0.74 at $t \simeq 35$ for $\Omega=1/2$. 
The difference is not large because the velocity has been
small prior to the nonlinear stage.  But it is the
Coriolis force that makes the rms variations saturated at a lower 
level and at an earlier time epoch in the case with rotation.
ii) The rms values of both $v_x$ and $v_z$ 
become converged in the case without rotation, while 
the rms of $v_z$ is always larger than that of $v_x$
in the case with rotation.  This is because the Coriolis force  
affects only the horizontal flow. 

\subsection{Energetics}

To examine the global properties of the instability further, we have followed
the evolution of kinetic, magnetic, gravitational, and heat
energies, which are defined as
\begin{equation}
E_k = \int\!\int\!\!\int \frac{1}{2}\rho (v_x^2+v_y^2+v_z^2) \, dxdydz,
\end{equation}
\begin{equation}
E_m = \int\!\int\!\!\int \frac{1}{8\pi} (B_x^2+B_y^2+B_z^2) \, dxdydz,
\end{equation}
\begin{equation}
E_g = \int\!\int\!\!\int \rho\;\phi\, dxdydz,
\end{equation}
\begin{equation}
E_h = \int\!\int\!\!\int\;p\;\ln p\, dxdydz,
\end{equation}
where integrals cover the whole computation box, $\phi=gz$ is the 
gravitational potential, and $p=\rho a^2$ is the gas pressure (Mouschovias 
1974). We take $\rho_0(0) a^2 H^3$ as the normalization unit for 
the energies.

Time evolution of the normalized
energies is plotted in Fig. 4a by solid lines for $\Omega=0$
and dotted lines for $\Omega=1/2$.  The general behavior
has a distinct feature at each evolutionary stage. 
The energies remain unchanged during the linear stage, show significant 
variation during the nonlinear stage, and then slowly converge to final 
values in the relaxed stage. During the linear stage, velocity is less
than the isothermal speed, and the Coriolis force is relatively small.
So we cannot notice any visible difference in energies between the two
cases with and without rotation.  However, as the Coriolis force 
becomes stronger in the nonlinear stage, differences appear.
Variation for $\Omega=1/2$ is less severe, showing a slower increase or
decrease than that for $\Omega=0$.  This is because the flow induced
by the instability has been slowed down by the Coriolis force.  But
the increase or decrease of energies lasts longer.  Hence, the energies
eventually converge to the same final values in both cases
in the relaxed stage.  This implies the final energetics of the instability 
is insensitive to rotation.

In Fig. 4a, we see a bump in the curve of the kinetic energy, $E_k$.
It is barely noticeable, although the rms 
velocities are comparable to the isothermal speed
during the nonlinear stage (see Fig. 3).
This is because most flow activities either occur in the upper, low
density regions or are localized in the magnetic valleys.  As gas in
higher positions accumulates into the magnetic valleys, gravitational
energy, $E_g$, is released and becomes available to the system.  Some of it
is used to increase the kinetic energy, and other is converted to work done 
by gas.  The heat energy, $E_h$, measures the amount of this work
(Mouschovias 1974).  The total magnetic flux is conserved in our
simulations due to the imposed boundaries, but the magnetic energy, 
$E_m$, decreases.  The reason is the following.  As the instability develops,
although it becomes compressed in a localized regions of
magnetic valleys, the magnetic field, which is initially 
concentrated around the equatorial plane (see Eq. 4), spreads upwards overall.
This is because magnetic field lines manage to slip away mass due to enhanced
reconnection.  A given amount of magnetic flux that is less concentrated 
has less energy. 
The final, converged value of the magnetic energy will be the
energy of uniformly distributed magnetic field whose flux
is same as the initial flux.

Mouschovias (1974) defined the energy integral of  
the system described in \S 2.1 as 
\begin{equation}
E_t = E_k + E_m + E_g + E_h,
\end{equation}
and showed that it is a constant of motion.  That is,
the energy released from certain forms should end up to 
increase the energy in other forms.  
Matsumoto et al. (1990) showed, however, that the energy integral
in their simulations of the Parker instability in a point-mass-dominated
gravity is not a conserved quantity when shocks are generated.
Our simulations have confirmed this.
The curves labeled with $E_t$ in Fig. 4b are constant up to the end of
the linear stage, since no shock has been developed yet.
However, as soon as the nonlinear stage begins (so the increase in
the kinetic energy becomes noticeable), shocks form in the valley
regions, so the energy integral starts to decrease.
When the system enters
the relaxed stage, it gradually settles to the final 
equilibrium state where all shocks have disappeared.
Then, the energy integral stays constant again.   

\subsection{Structures}

To illustrate the structures resulting from the instability,
three-dimensional images are shown in Fig. 5.  The perspective images
of density structure and magnetic field lines are at two epochs of the 
linear stage ($t=30$, upper panels) and the nonlinear stage ($t=36$,
lower panels), respectively.  The images only for $\Omega=1/2$ are presented,
since the corresponding images for $\Omega=0$ look similar.
Also refer to the images for $\Omega=0$ in Fig. 2 and in Kim et al. (1998),
although they are drawn at different epochs.
A gray surface with density $\rho_0(z=4)$ is depicted in each left panel
as in Fig. 2, while 64 field lines, whose starting points lie along the
line of $z=4$ and $y=0$, are drawn in each right panel.  At $t=0$,
the field lines are straight and lie in the $z=4$ plane.

As mentioned in \S 2.2, in the linear stage, the interchange mode induces
alternating regions of compression and rarefaction along the
$x$-direction, while the undular mode bends magnetic field lines vertically.
Those features are clearly seen in the images of the linear stage.
The isodensity surface in the left upper panel of Fig. 5 is corrugated along the $x$-direction, 
which is the defining feature of the interchange mode.  
The field lines in the right upper panel go up and down along the $y$-direction
making valley and arch regions,
which is the distinguishing feature of undular mode.  
As the instability develops, the alternating structures become
complicated.  At the same time, the density contrast between valley and
arch regions becomes large.  The isodensity surface in the left lower panel
shows bumps at the valley and dents at the arch regions.  
There is no regular occurrence pattern of the bumps and dents, even
though the corrugations of the surface along 
the $x$-direction are still preserved in this stage.  
The distribution of the field lines becomes complicated too.
As more gas moves along field lines from the 
arch to the valley, the arch field lines move to even higher positions
and the valley field lines move down to lower positions,
as clearly shown in the right lower panel.  

An effect of rotation on nonlinear structures is illustrated
in Fig. 6, where two field lines
together with velocity vectors along them are plotted at $t=40$.
Both are examples of valley and arch field lines.  Flow velocity
along those undulated field lines becomes large in the
nonlinear stage.  So flows are under the strong influence 
of the Coriolis force.  As gas in-falls along the valley field line,
flow velocity increases and reaches its maximum around the bottom of the valley.
The near side gas along the left field line in Fig. 6 experiences the force
in the $+x$ direction, while the far side gas experiences the oppositely
directed force.  These forces induce the largest oppositely twisting
velocities around the bottom, which cause the bottom part of the undulated
field line to become twisted.  A similar situation occurs along the arch field
line in the right side of Fig. 6.  This time, the near side gas from
the arch is deflected toward the $-x$ direction and the far side gas
is deflected toward the $+x$ direction.
However, the flow velocity is smaller around
the tip of the arch, so the field line is not twisted.  Comparing the
field lines at the valley and arch regions, we find that
the curvature of the arch field is larger than that of the valley field.

Note we have started our simulations only with uniform magnetic field.
But the field lines, which have been bent in magnetic valley regions, get
twisted due to the Coriolis force.  Since the direction of twisted 
lines changes rapidly, they may be regarded as a ``random'' field.  So the 
Coriolis force has played a role
of converting the uniform component of magnetic
field into the random one.  Assuming that the scale of the twisting is
comparable to the size of condensations formed in valley regions, 100 pc
(see the next subsection), it is not far from the single-cell-size of
the Galactic random field measured in Rand and Kulkarni (1989).  

\subsection{Column Density}

The structures formed are further illustrated by examining the
column densities defined by
\begin{equation}
N_z(x,y;t) = {\int_{0}^{12} \rho(x,y,z;t) \, dz \over 
\int_{0}^{12} \rho(x,y,z;0) \, dz}_,
\end{equation}
\begin{equation}
N_x(y,z;t) = {\int_{0}^{12} \rho(x,y,z;t) \, dx \over 
\int_{0}^{12} \rho(x,y,0;0) \, dx}_,
\end{equation}
\begin{equation}
N_y(x,z;t) = {\int_{0}^{12} \rho(x,y,z;t) \, dy \over 
\int_{0}^{12} \rho(x,y,0;0) \, dy}_. 
\end{equation}
We have kept track of the evolution of
the vertical column density, $N_z$, and plotted in Fig. 7
its minimum and maximum as functions of time.
Solid lines correspond to $\Omega=0$ and dotted lines to $\Omega=1/2$.
Both cases show a similar trend of evolution.  Initially
$N_z$ stays close to unity for all $x$ and $y$.
As gas starts to fall into valleys
in the linear stage, the contrast between the minimum and
maximum increases.  It reaches peaks in the nonlinear stage and
decreases during later evolution.  In addition, there are differences between
the two cases:  i) the peak of the maximum $N_z$
for $\Omega=1/2$ (1.5) is smaller than that (2.1) for $\Omega=0$,
indicating the nonlinear structures are less acute, ii) the maximum $N_z$ for
$\Omega=1/2$ peaks at later time, indicating the nonlinear structures
develop more slowly, iii) the maximum $N_z$ for $\Omega=1/2$
is broader, indicating the nonlinear activity lasts longer.  All these
are attributed to the effects of the Coriolis force, and agree with
the descriptions in \S 3.2 and 3.3.

At $t=35$ for $\Omega=0$ and $t=42$ for $\Omega=1/2$ around the peaks of
the maximum $N_z$, the distributions of $N_z$ in the $x-y$ plane are
drawn in Fig. 8.  In each panel, equal 
column-density contours overlie a gray-scale map.  Thick solid 
lines that correspond to $N_z=1$ trace out the
boundaries of over and under-dense regions.
Over-dense regions are labeled with thin solid lines and dark grays, whereas 
under-dense regions are labeled with dotted lines and light grays.  In
the case without rotation (see Fig. 8a) sheet-like structures
form, whose long dimension is aligned with the initial field ($y$-) 
direction, as pointed out in Kim et al. (1998).
Similar structures are formed in the case with rotation,
but the long dimension is a bit tilted with respect to the $y$-direction 
and the shape is a bit rounder (see Fig. 8b).
The tilting is of course the consequence of the Coriolis force,
$-2 \Omega {\hat z} \times {\bf v}$.
Note that due to the periodic nature of our boundaries, the allowed
angle for tilting is quantized and given by
\begin{equation}
\theta_t \equiv \arctan\left(k_x \over k_y \right),
\end{equation}
where $k$'s can have the values of $2\pi/12H$, $2(2\pi/12H)$, $\cdots$,
$256(2\pi/12H)$ ($12H$ is the box size).
Fig. 8b shows that $\theta_t$ is close to $\arctan(1/6)$,
corresponding to $k_x = 6(2\pi/12H)$ and $k_y = 2\pi/12H$. Those coincide
with ($k_x$, $k_y$) of the structure most visible in the figure.

While the $N_z$ map illustrates the vertically
developed structures, $N_x$ and $N_y$ are more closely compared with
observations in the Galaxy.
Considering that the initial field direction is parallel to $y$-direction 
and the sheet-like structure is almost aligned with that direction, $N_x$ and 
$N_y$ represent two extremes: the line-of-sight is perpendicular 
($N_x$) to and parallel ($N_y$) to the mean field direction.
The observed column 
density would be somewhere between the two.  Contours with equal $N_x$ and
$N_y$ are plotted in left and right panels of Fig. 9, respectively, 
for the case with rotation at $t=36$.  We have chosen 10 
contours whose column densities are $e^{-1}, e^{-2}, \cdots, e^{-10}$ of 
the initial midplane value.  So initially the contours were straight
and equally spaced.  Each contour in the left panel shows one or two bumps
along the $y$-direction,
which are attributed to the manifestation of the undular mode.
On the other hand, the counterpart in the right panel is more irregular and
shows many bumps along the $x$-direction, which are attributed to
the manifestation of the interchange mode.

\section{SUMMARY AND DISCUSSION}

We have studied the evolution of the Parker instability in non-rotating
and uniformly rotating magnetized disks through high resolution numerical 
simulations (using up to $256^3$ grid zones).  The main purpose of this 
paper is to address the effects of rotation on the instability by
comparing the two cases.  Here our findings are summarized.

First, the overall evolution of the instability in a uniformly
rotating disk is similar to that in a non-rotating disk.  As in the case
without rotation (Kim et al. 1998), the evolution in the case with
rotation is divided into three stages: linear, nonlinear, and relaxed.
In the linear stage, the mixed mode regulates the evolution, as
predicted in linear stability analyses.  That is, exponentially
growing perturbations have a preferred scale along the azimuthal direction
but have the smallest possible scale along the radial direction.
In the nonlinear stage, the growth is saturated.
The undular mode bends magnetic field lines, forming valleys and arches,
but overall flow motions appear chaotic. 
After becoming fully developed,
the nonlinear oscillatory structures damp out, and at the
same time, reconnection enhanced by chaotic motions allows
re-distribution of mass and magnetic field. 
At this stage, the system
enters the relaxed stage, although the boundary between the nonlinear
and relaxed stages is not as clear as in the case without rotation.

Second, the Coriolis force introduces a couple of visible differences.
One of them is the twisting of magnetic field lines in valley regions.
In the case with rotation, gas is subject to the Coriolis force,
whenever there are lateral motions.
But the force is especially strong in valley regions,
because the falling speed of gas toward those regions becomes comparable to 
or faster than the isothermal sound speed.  Flows approaching valleys
from two sides along the field lines experience the oppositely directed
Coriolis forces, so the field lines there get twisted.
The scale of the twisting is comparable to the size of the
valleys, and is $\sim100$ pc or less.  It is noticeable that this scale of
twisting is compatible with ``single-cell size'' of random magnetic field
measured by Rand \& Kulkarni (1989).

Third, we have found that the maximum enhancement factors of vertical column 
density are 2.1 for the case without rotation, and 1.5 for the case with 
rotation.  In the former case, the enhanced sheet-like structures, viewed 
from the vertical column density map, are aligned with the direction of
mean magnetic field.  On the other hand, in the latter case, the
structures are slightly tilted with respect to the field direction and a bit
rounder.  This is because the lateral motions of gas are affected by the 
Coriolis force.

In our previous study of Kim et al. (1998),
we concluded that it is difficult to regard the Parker instability
{\it alone} as the formation mechanism of GMCs, because
i) the sheet-like shape of the structures formed is different from the
shape of GMCs and ii) the density enhancement factor
is too small. The inclusion of rotation does not change this conclusion.  
However, we do not completely rule out the role of the Parker instability
in the formation of GMCs.
Inclusion of radiative cooling and self-gravity may result
in thermal and gravitational instabilities, which certainly increase  
the density enhancement factor far more than the one brought by the Parker
instability alone and shape structures in a different way.  In the context
of a more complicated system with cooling and self-gravity, the Parker
instability agglomerates diffuse gas into magnetic valley regions, which 
provide a more conducive environment to both thermal and gravitational
instabilities.  Afterward, they may override the Parker instability, 
especially in each magnetic valley region, and govern later evolution.   
Here we want to emphasize that other physical processes, such as 
thermal and gravitational instabilities, are required to get
more than factor of 2 increase in vertical column density.    

Because of the sheet-like shape of the nonlinear structures aligned with
the mean magnetic field, the horizontal column density, which is an
observable quantity in the Galaxy, depends on the angle between the
line-of-sight and the field direction.  When they are aligned 
with each other, the column density map illustrated in the right panel
of  Fig. 9 shows vertical filamentary structures. So we suggest that
the Parker instability would be a possible mechanism for the formation
of the interstellar filamentary structures.
The ``filament 1'' revealed in a H$\alpha$ survey (Haffner, Reynolds, \& 
Tufte 1998) is the most plausible observed candidate that may have
been induced by the Parker instability.   It is about kpc-long, has
arc-shaped morphology, and resides at high latitude.
These are essentially the attributes of the Parker instability.
It is, however, difficult to make a direct comparison 
between our simulations and their observations, because
the Galactic environment at the location (about 1 kpc away from
the sun to the direction of $l = 225^{\circ}$) of the ``filament 1'' 
could be different from that in the solar neighborhood. It may be an
interesting project to perform a direct numerical modeling for
the filament 1.

In this paper, we have studied the effects of uniform rotation, although 
differential rotation is a proper description of the galactic rotation 
law.  Inclusion of differential rotation into magnetized gas disks would
complicate the situation, because not only the Parker instability but also
the magneto-rotational instability (MRI; Balbus \& Hawley 1991; Hawley \& 
Balbus 1991) can come into existence.  The evolution of the two instabilities
in accretion disks was studied, for instance, by Miller \& Stone (2000)
through three-dimensional simulations.  However, the MRI in the galactic disk
is not as active as in accretion disks or even can be stabilized,
because the galactic magnetic field is very strong.
But it is not clear whether the MRI is completely suppressed
for the following reason.  As the Parker instability
proceeds,  high-$\alpha$ arch regions and low-$\alpha$ valley regions
are segregated.  The arch regions are stable to the MRI, due to even
stronger magnetic fields, but the valley regions may be subject to the MRI.  
The action of the two instabilities on magnetic field
is different. The Parker instability pumps field lines vertically, 
while the MRI stretches them radially.  Together these
instabilities and differential rotation,
along with dissipation and/or reconnection mechanisms of magnetic 
field may serve as the ingredients of the galactic dynamo 
(Tout \& Pringle 1992).  This line of approach on the galactic dynamo 
would be worthy of further detailed investigations.

\acknowledgments

We would like to thank Don Cox and Bob Benjamin, who reviewed this paper.  
They picked up a mistake in Fig. 4 and provided several suggestions/comments 
which improved the paper.
The work by DR was supported in part by the grant KRF-99-015-DI0113.
The work by TWJ was supported in part by the NSF through grants
INT95-11654 and AST96-19438, by NASA grant NAG5-5055 and by the
University of Minnesota Supercomputing Institute.

\clearpage

\begin{deluxetable}{ccccc} 
\tablehead{
\colhead{Model\tablenotemark{a}} & 
\colhead{$N_x \times N_y \times N_z$} & 
\colhead{computational domain\tablenotemark{{b}}} &
\colhead{$t_{\rm end}$\tablenotemark{c}} & 
\colhead{$\Omega$\tablenotemark{c}}
}  
\startdata
1    & $~64 \times ~64 \times ~64$ & $0 \le x,y,z \le 12$ & 100 & 0   \\
2    & $128 \times 128 \times 128$ & $0 \le x,y,z \le 12$ & 100 & 0   \\
3    & $256 \times 256 \times 256$ & $0 \le x,y,z \le 12$ &  93 & 0   \\
4    & $256 \times 256 \times 256$ & $0 \le x,y,z \le 12$ &  70 & 1/2 \\
\enddata
\tablenotetext{a}{All models have $\alpha=1$, initially.}
\tablenotetext{b}{The length unit is the scale height $H$.} 
\tablenotetext{c}{The time unit is the sound travel time over the
scale height $H/a$.}

\end{deluxetable}

\clearpage

\begin{figure}
\label{fig:DR}
\epsfxsize=14truecm
\centerline{\epsfbox{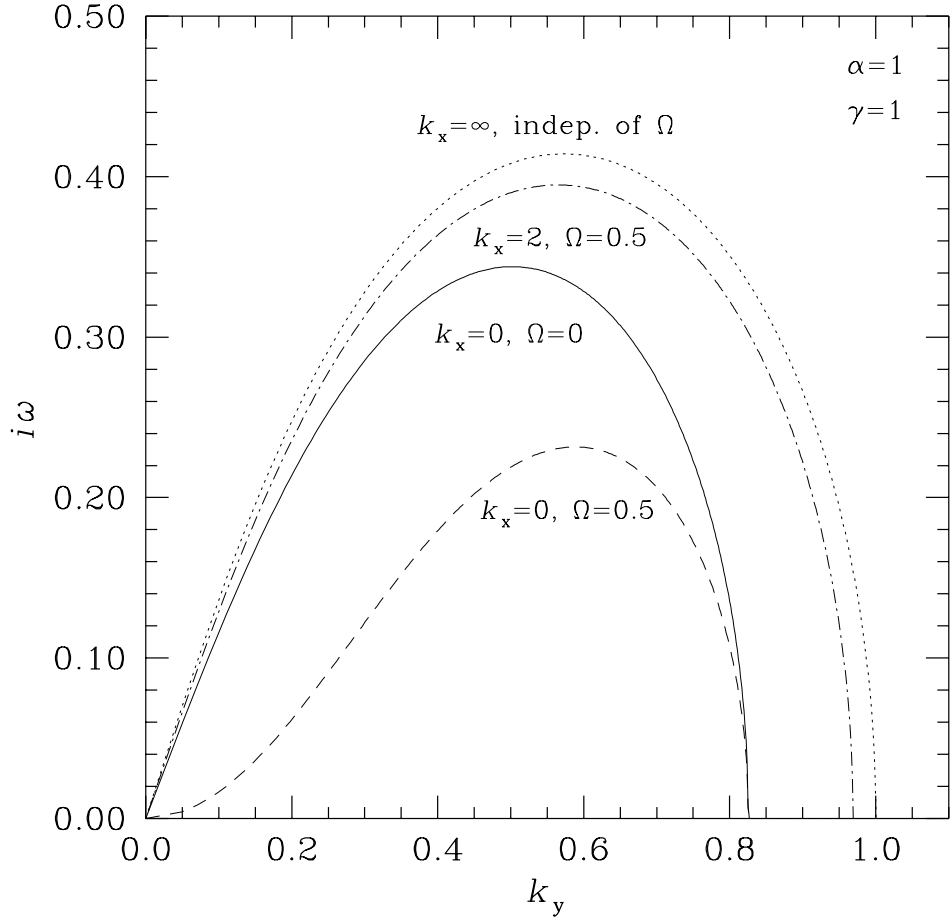}}
\figcaption{Dispersion relations for the Parker instability in exponentially
stratified magnetized disks with and without rotation.  The ordinate 
represents the growth rate and the abscissa represents the azimuthal 
wavenumber. Each curve is marked by radial wavenumber and angular speed.
The normalized units for the wavenumbers, the growth rate and 
the angular speed are $1/H$, $a/H$, and $a/H$, 
respectively, where $H$ is the scale height and $a$ is the isothermal speed.}
\end{figure}

\begin{figure}
\label{fig:resol}
\vspace{-5truein}
\epsfxsize=24truecm
\centerline{\epsfbox{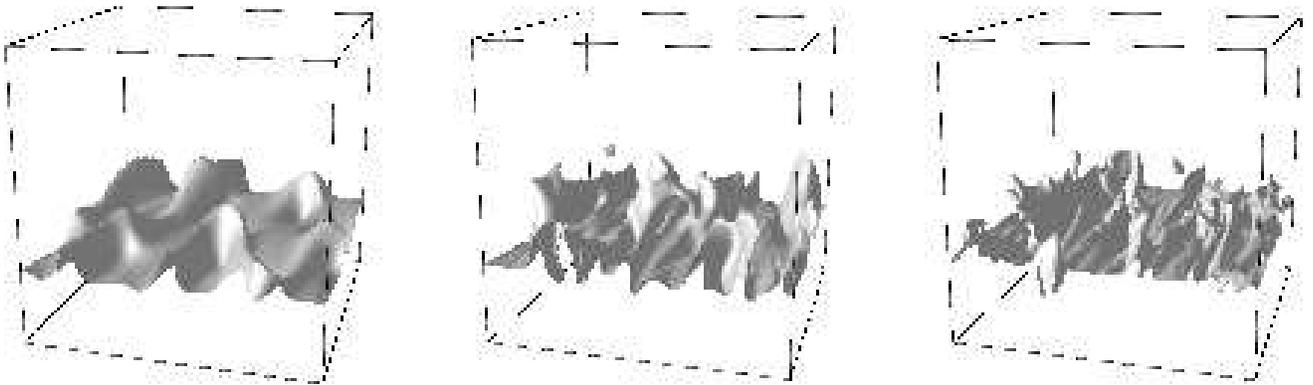}}
\figcaption{Comparison of density structure simulated with different number
of grid zones: a) coarse ($64^3$, Model 1), b) medium ($128^3$ Model 2),
and c) high ($256^3$, Model 3).
The computational box is oriented in such a way that the radial ($x$), 
the azimuthal ($y$), and the vertical ($z$) directions are from left to 
right, from near to far, and from bottom to top, respectively.
At the same epoch $t=35$, gray surface with density $\rho_0(z=4)$ is shown.
The unit of time is $H/a$, and the size of the computational
box is $(12H)^3$.}
\end{figure}

\begin{figure}
\label{fig:vrms}
\epsfxsize=19truecm
\centerline{\epsfbox{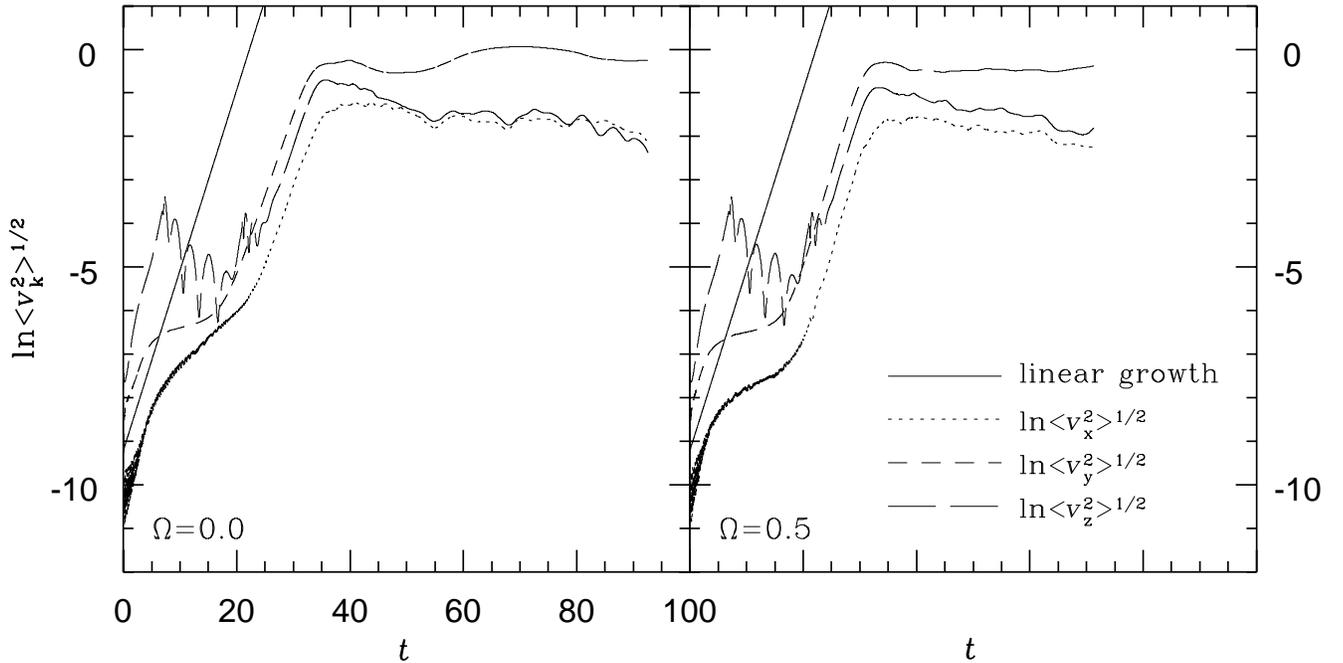}}
\figcaption{Rms of each velocity component as a function of time.
The left panel is for the case without rotation and the right panel
is for the case  with rotation.  Natural log is used along the ordinate.    
The maximum linear growth rate, 0.41, is represented by the solid line.
The units of velocity and time are $a$ and $H/a$, respectively.}
\end{figure}

\begin{figure}
\label{fig:E}
\epsfxsize=19truecm
\centerline{\epsfbox{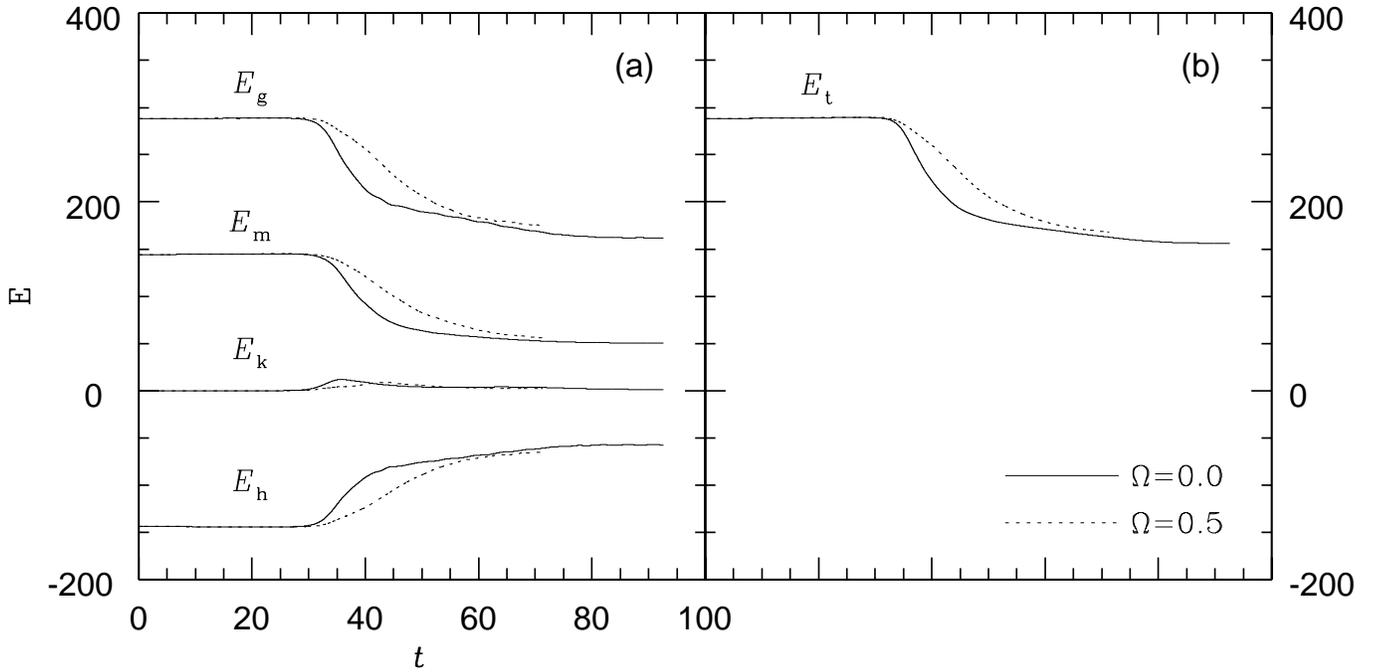}}
\figcaption{Evolution of (a) kinetic $E_k$, magnetic $E_m$, 
gravitational $E_g$, and heat energies $E_h$, and (b) energy integral 
$E_t$ ($=E_k+E_m+E_g+E_h$).  Energies for the case without rotation are
represented by solid lines, whereas those with rotation are by dotted lines.
The units of energy and time are $\rho_0(0)a^2H^3$ and $H/a$, respectively.}
\end{figure}

\begin{figure}
\label{fig:rhoB}
\vspace{-1.5truein}
\epsfxsize=18truecm
\centerline{\epsfbox{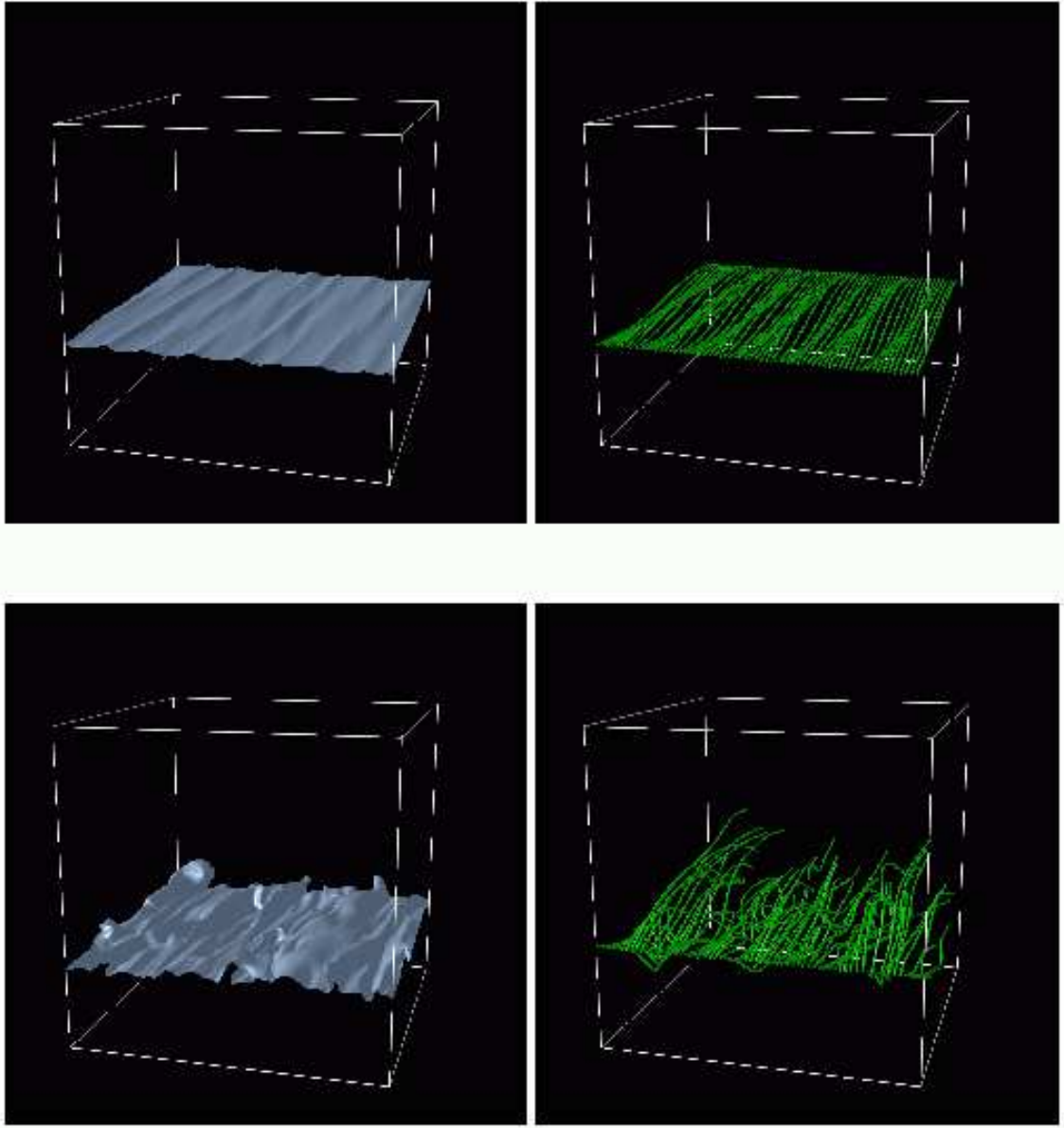}}
\figcaption{Perspective view of density structure and magnetic field
lines at the epochs $t=30$ (top) and $t=36$ (bottom) from the
high resolution simulation with rotation (Model 4).  The computational boxes 
are oriented in the same way as in Fig. 2.  Gray surface with isodensity
$\rho_0(z=4)$ is shown in the left panels. 
Magnetic field lines (64), whose starting points lie along the line of
$z=4$ and $y=0$, are represented by green tubes in the right panels.
The units of length and time are $H$ and $H/a$, respectively.}
\end{figure}

\begin{figure}
\label{fig:BV}
\epsfxsize=10truecm
\centerline{\epsfbox{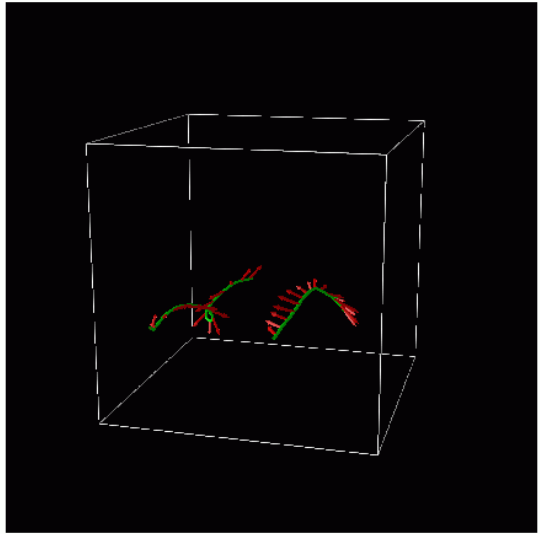}}
\figcaption{Two magnetic field lines and velocity vectors along them at the
epoch $t=40$ from the high resolution simulation with rotation (Model 4).
The computational box is oriented in the same way as in Fig. 2.
Field lines are drawn with green, and velocity vectors are drawn with red.
Same normalization units in Fig. 5 are used.}
\end{figure}

\begin{figure}
\label{fig:Nminmax}
\epsfxsize=18truecm
\centerline{\epsfbox{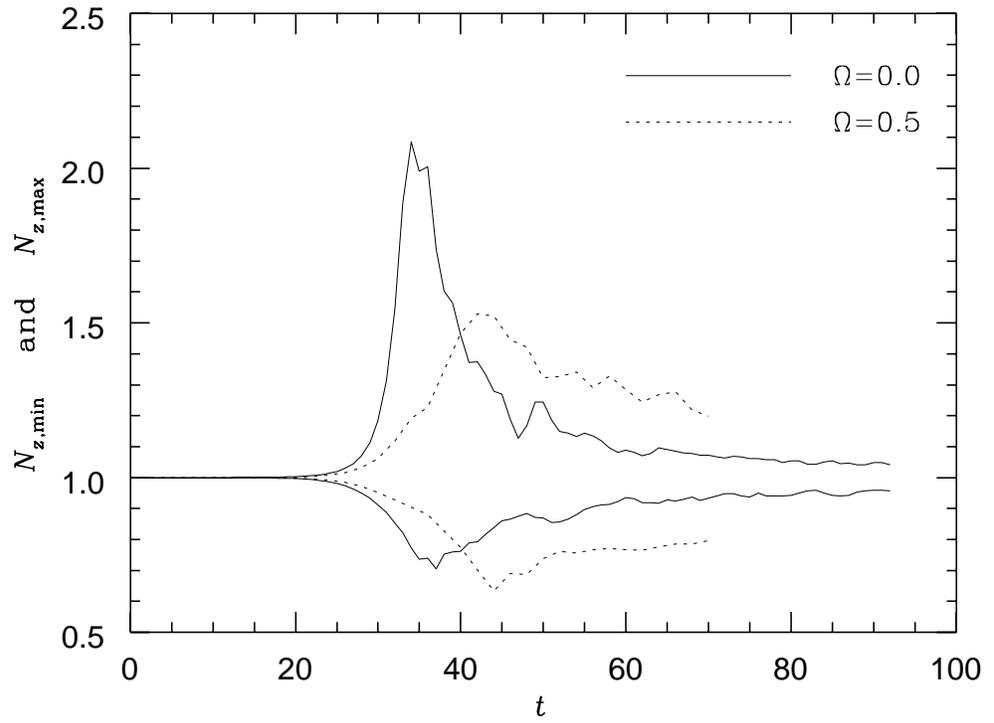}}
\figcaption{Maximum and minimum of the vertical column densities as 
functions of time.  The column density has been normalized to its initial 
value.  The solid lines represent the case without rotation,
and the dotted ones the case with rotation.  The unit of time is $H/a$.}
\end{figure}

\begin{figure}
\label{fig:Nz}
\epsfxsize=15truecm
\centerline{\epsfbox{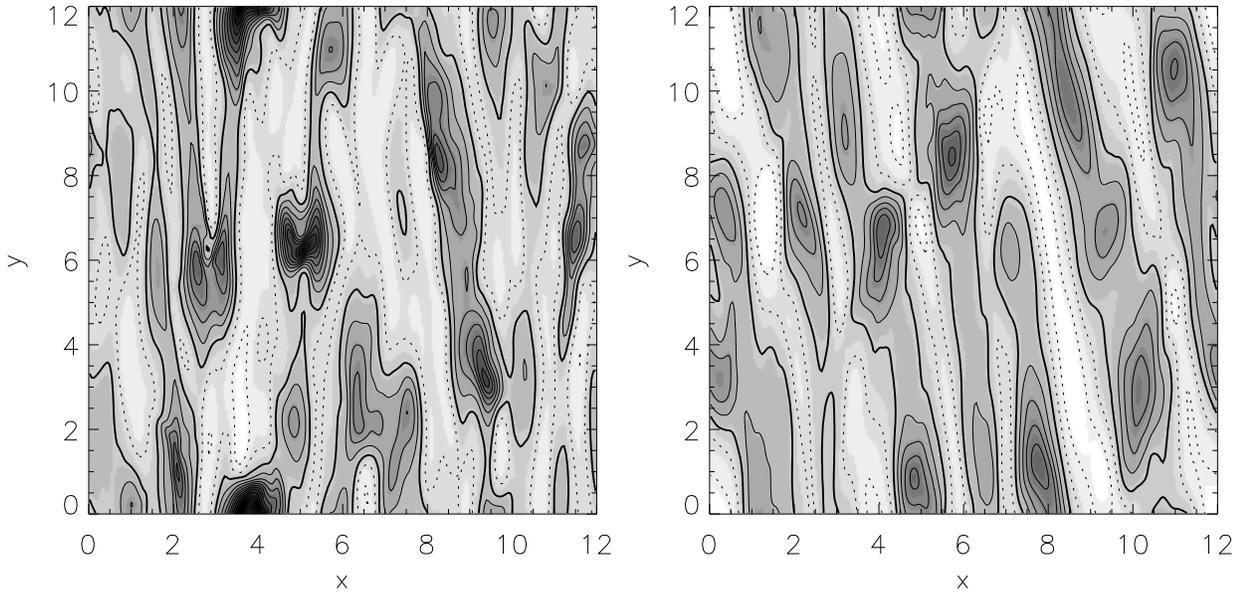}}
\vspace{1truein}
\figcaption{Vertical column density maps for the cases (a) without
rotation at $t$=35 (b) with rotation at $t$=42.  The time epochs have been
chosen so that the column density reaches peaks at those epochs. 
The gray scale image of the column density is overlaid with equal
density contours.
Solid and dotted lines represent over-dense and low-dense regions,
respectively, which are divided by thick solid lines.
The interval is 1/10 of the initial value.
The units of length and time are $H$ and $H/a$, respectively.}
 \end{figure}

\begin{figure}
\label{fig:Nxy}
\epsfxsize=15truecm
\centerline{\epsfbox{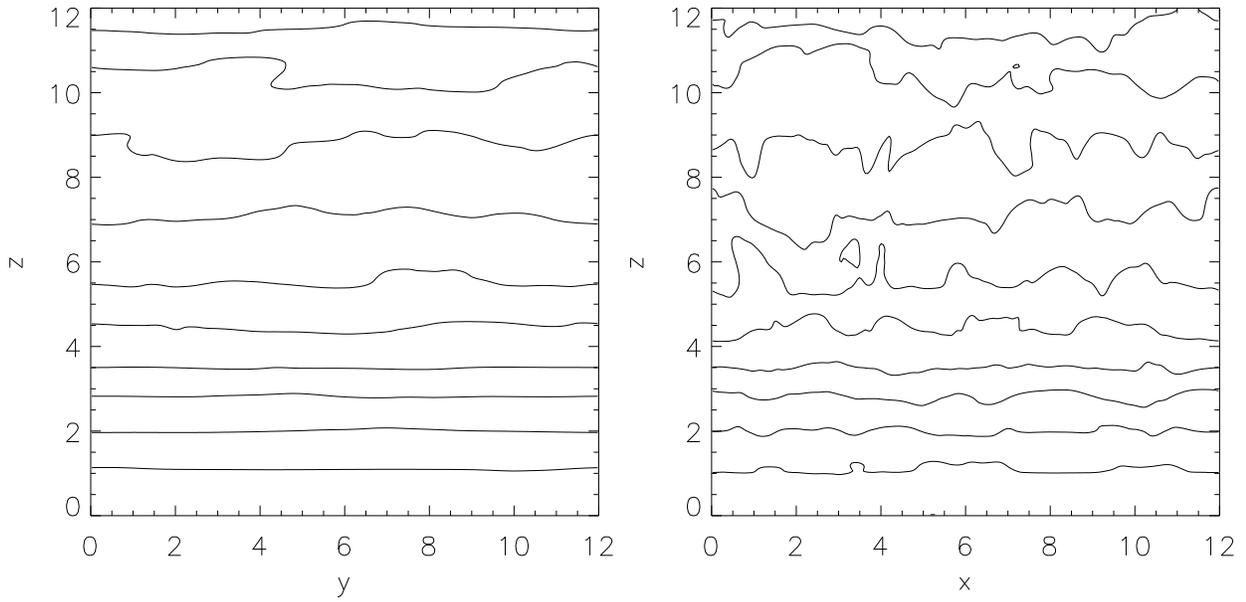}}
\vspace{1truein}
\figcaption{Horizontal column density maps for the case with rotation
at $t=36$.  Density has been integrated along the radial direction in
the left panel and along the azimuthal (mean magnetic field) direction
in the right panel.  10 equal column-density contours represent the lines
with $e^{-1}, e^{-2}, \cdots, e^{-10}$ of the initial midplane value.
The units of length and time are $H$ and $H/a$, respectively.}
\end{figure}

\end{document}